\documentclass[11pt]{article}
\usepackage{graphics}
\usepackage{epsfig}

\newcommand{\be}{\begin{eqnarray}}
\newcommand{\ee}{\end{eqnarray}}

\def\ll#1{\left#1}
\def\r#1{\right#1}
\def\fr{\frac{1}{2}}

\def\mref#1{(\ref{#1})}

\def\p{\partial}

\def\bd{\begin{displaymath}}
\def\ed{\end{displaymath}}

\def\ba#1{\begin{array}{#1}}
\def\ea{\end{array}}
\def\nn{\nonumber}
\newfont{\Bbb}{msbm10 scaled 1200}

\newtheorem{tw}{THEOREM}
\newtheorem{de}{DEFINITION}

\begin{document}

\pagestyle{empty}

\begin{center}

{\LARGE\bf High energy semiclassical wave functions in the Bunimovich stadium billiards determined by its periodic orbits\\[0.5cm]}

\vskip 12pt

{\large {\bf Stefan Giller}}

\vskip 3pt

Jan D{\l}ugosz University in Czestochowa\\
Institute of Physics\\
Armii Krajowej 13/15, 42-200 Czestochowa, Poland\\
e-mail: stefan.giller@ajd.czest.pl
\end{center}

\vspace {40pt}

\begin{flushright}
{\it Mojej \.Zonie}
\end{flushright}

\vspace{20pt}

\begin{abstract}
It is argued that the high energy semiclassical wave functions (SWF) in an arbitrary billiards can be built by approximating the billiards by a respective
polygon one. The latter billiards is determined by a finite number of periodic orbits of the original one limited by their lengths beginning with the
shortest ones and which are common for both the billiards. The phenomenon of scars and superscars (Heller, E.J., {\it Phys. Rev. Lett.} {\bf 53},
(1984) 1515) are then naturally incorporated into such a construction being a limit of periodic orbit channels (POCs) considered by Bogomolny and Schmit
({\it Phys. Rev. Lett.} {\bf 92} (2004) 244102). The Bunimovich stadium billiards is considered as an example of such an approach.
\end{abstract}

\vskip 50pt
\begin{tabular}{l}
{\small PACS number(s): 03.65.-w, 03.65.Sq, 02.30.Jr, 02.30.Lt, 02.30.Mv} \\[1mm]
{\small chaotic dynamics, quantum chaos, scars, superscars}
\end{tabular}

\newpage

\pagestyle{plain}

\setcounter{page}{1}

\section{Introduction}

\hskip+2em Since the well known Gutzwiller papers (see \cite{3} for references to these papers) we have been learned about the fundamental role played by periodic orbits of the classical
motions in obtaining the semiclassical limits of respective quantum problems. However this fundamental role mentioned is visible so far only in the Feynman path
formulation of the quantum mechanics as it was shown by Gutzwiller and is completely absent as it seems in the wave function formalism of the quantum
mechanics used by Maslov {\it et al} \cite{4} to get the semiclassical limit of the theory. In fact in the latter case it appears that only these quantum
system can be treated unambiguously by the method the classical limits of which are integrable. This is in a contrast with the Gutzwiller approach which can
applied also to the quantum systems which classical limit is chaotic.

Recently however studying the polygon billiards we have shown
\cite{41}-\cite{42} that the Maslov approach can be extended also to pseudointegrable system of classical billiards although with some restrictions.
Nevertheless
still in this extended applications the classical periodic orbits if present seem do not play some distinguished role just providing only examples of
specific states of the quantized polygons called superscars \cite{53}-\cite{44}.

Considering however the polygon billiards and the billiards with some arbitrary forms of their boundaries it is a natural temptation to approximate the
boundaries of the latter by the ones of the former to apply to the obtained polygons the semiclassical quantization method worked out for the polygon
billiards hoping for some satisfactory approximation of the quantities corresponding to the original billiards. Such expectations are relied on the known
theorems about dependence of the eigenvalues on continuous changes of billiards boundary, see \cite{40} and App.C.

A problem which however immediately appears with
such an approach to an arbitrary billiards is of course that there is a priori an infinite number of possible different polygon billiards by which such an
arbitrary one can be approximated. Therefore we should have some criterion which
\begin{itemize}
\item allows us for a unique choice of a definite polygon billiards from an infinite set of them approximating the considered one;
\item provides us with the polygon billiards which are related to the original one in some inherent way; and
\item allows us for controlling of levels of approximations of the original billiards by respective polygon ones.
\end{itemize}

Such a criterion can be provided just by periodic orbits of the considered non-polygon billiards. Namely it seems to be obvious that knowing all periodic
orbits of a billiards we should be able to recover its boundary fully and uniquely. Such a convince which appeals to the Poincare recurrence theorem
(see for example \cite{5}) and to continuity arguments allows us to claim that a set of all
points of the billiards boundary which the periodic trajectories is reflected off is {\it dense} on the boundary. Of course since the set of periodic
orbits is in general unknown fully we have to limit ourselves to a number of them which can be found in this or other way also numerically. One
can limit a set $P_L$ of them just by limiting their maximal lengths to a (real) number $L$ which otherwise can be fixed arbitrarily. Of course the
larger $L$ the larger is the set $P_L$ of periodic orbits which enter it and the closer each other are the boundary reflection points of
these periodic orbits.

Now if $L$ is fixed and the respective set $P_L$ is known then in each point of the billiards boundary reflecting any periodic orbit of $P_L$ a tangent
to the boundary can be drawn. Extending each such a tangent to cross it with the closest two neighbour ones we construct in a unique way a polygon for which the orbits of
$P_L$ are the subset of all the periodic orbits of the polygon obtained in this way. This is just the idea which will be used in the paper to construct
approximately semiclassical wave functions in billiards with arbitrary boundaries.

In general the polygon billiards approximating the original one in the above way are irrational, i.e. their angles measured in the $\pi$-unit are
irrational. Nevertheless to tackle this problem we can take into account our earlier paper where the problem of building of the high energy SWFs in the
irrational polygon billiards has been discussed \cite{53}-\cite{44}.

In sec.2 we apply the approach described above to the well known billiards with the chaotic non-integrable classical motion which the
Bunimovich stadia are \cite{52}. Three variants of the stadia shown in Fig.1 are considered and the SWF is built for each case together with the corresponding energy
spectrum.

The remaining part of the paper is organized as follows.

In sec.3 the superscar phenomena appearing in the semiclassical approximations built for the considered Bunimovich stadia are discussed.

In sec.4 the accuracy of the semiclassical approximations provided by the approach described in this introduction is estimated.

In sec.5 an enveloping of the Bunimovich stadium by an irrational polygon billiards is considered.

Sec.6 is devoted to summarizing and discussing the results of the paper.

\section{Semiclassical wave functions built in the Bunimovich stadium billiards}

\hskip+2em Consider the Bunimovich stadia shown in Fig.1$A,B,C$. The semicircles of the stadia have the radius equal to one, while their flat parts have
the respective lengths equal to $2L$ shown in the figure. The lengths have been chosen to ensure the respective enveloping polygon billiards to be
the rational ones. An irrational case will be considered in sec.5.

The polygons $A',B',C'$ approximating the respective stadia have been built according to the method described in Introduction using the periodic
orbits of the respective stadia shown in Fig.1A,B,C. There are eight of such orbits in the cases $A$ and $C$ of the stadium and seven of them in the
case $B$.

As it was mentioned earlier due to the chosen lengths of the flat parts of the stadia and due to the periodic orbits chosen all the obtained polygon
billiards are rational. In the
case $A'$ of the polygon billiards all its sixteen angles are equal to $\frac{7}{8}\pi$. In the remaining two cases the respective polygons have twenty
angles each eleven of which are equal to $\frac{7}{8}\pi$ each while eight of them are equal to $\frac{15}{16}\pi$ each.

It is clear however that constructing SWFs in each of the polygon billiards mentioned it is enough to limit such constructions to a quarter of it and
extending the SWFs got in this way on the whole area of the billiards by the symmetry arguments. Therefore the respective constructions will be done in
the corresponding polygons shown in Fig.1$A'',B'',C''$. We shall consider their cases consecutively.

On the beginning let us estimate a common accuracy of the approximations provided by substituting the original stadia by the respective polygon
billiards. This can be done by constructing transformations of the Bunimovich stadia areas into the polygon ones to satisfy THEOREM 4 of App.C.
They can be following
\be
x'=\ll\{\ba{lrr}
        x&\;\;\;\;\;\;\;\;\;\;\;\;\;\;\;\;\;&-L\leq x\leq L\\
        \frac{x}{|x|}L+r(\phi)(x-\frac{x}{|x|}L)=x+(r(\phi)-1)(x-\frac{x}{|x|}L)&\;\;\;\;\;\;\;\;\;\;\;\;\;\;\;\;\;&L<|x|\leq L+1
        \ea\r.\nn\\
y'=\ll\{\ba{lr}
        y&-L\leq x\leq L\\
        r(\phi)y=y+(r(\phi)-1)y\;\;\;\;\;\;\;\;\;\;\;\;\;\;\;\;\;\;\;\;\;\;\;\;\;\;\;\;\;\;\;\;\;\;\;\;\;\;\;\;\;\;\;\;\;\;\;\;\;\;\;\;\;\;\;\;\;\;\;&L<|x|\leq L+1
        \ea\r. \nn\\
        -\fr\pi\leq\phi\leq\frac{3}{2}\pi
\label{2}
\ee
where $r(\phi)$ is shown in Fig.1$A'$.

\begin{figure}
\begin{center}
\psfig{figure=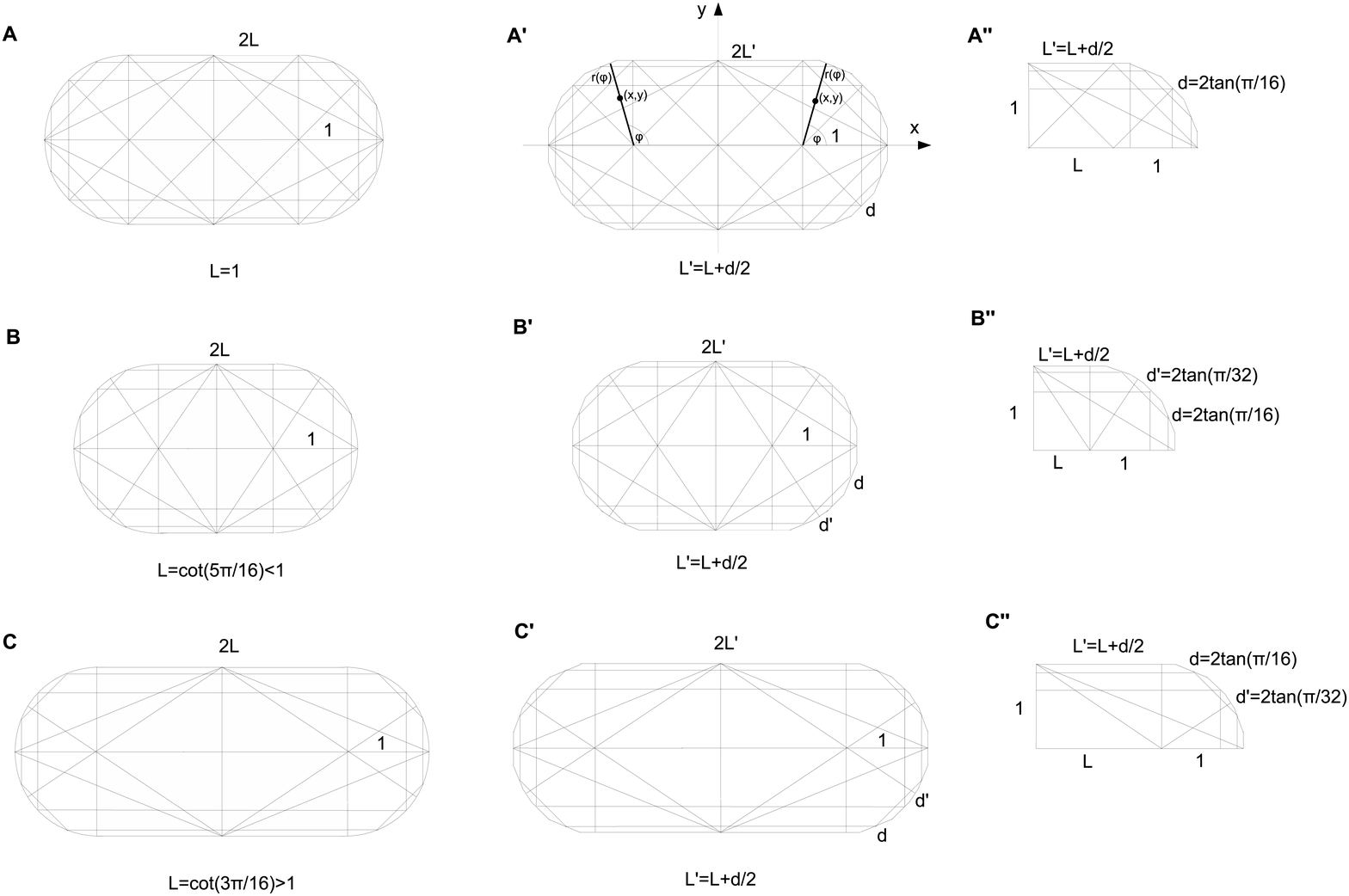,width=15cm} \caption{The Bunimovich stadia $A, B, C$ and their respective polygon approximations $A', B', C'$. The latter
are all rationals. The polygons $A'',B'',C''$ are the respective quarters of the $A', B', C'$ ones. The periodic orbits of the stadia used to construct
the approximating polygons which are shown in each of the billiards $A, B, C$ are unchanged in the respective polygons}
\end{center}
\end{figure}

The following estimations can be got easily from \mref{2}
\be
|(r(\phi)-1)(x-\frac{x}{|x|}L)|=|(r(\phi)-1)\cos\phi|\leq\frac{2\sin^2\frac{\pi}{32}}{\cos\frac{\pi}{16}}=0,0196=\epsilon_{Pol}\nn\\
|(r(\phi)-1)y|=|(r(\phi)-1)\sin\phi|\leq\epsilon_{Pol}\nn\\
-\fr\pi\leq\phi\leq\frac{3}{2}\pi
\label{2a}
\ee
so that the energy levels $E_n$ of the Bunimovich stadia $A,B,C$ of Fig.1 are approximated by the correspondingly ordered energy levels $E_n^{pol}$ of their
polygon envelopes $A',B',C'$ with an accuracy $\eta_{pol}$
\be
\ll|\frac{E_n^{pol}}{E_n}-1\r|<\eta_{pol}
\label{2b}
\ee
where $\eta_{pol}$ depends only on $\epsilon_{pol}$.

\begin{figure}
\begin{center}
\psfig{figure=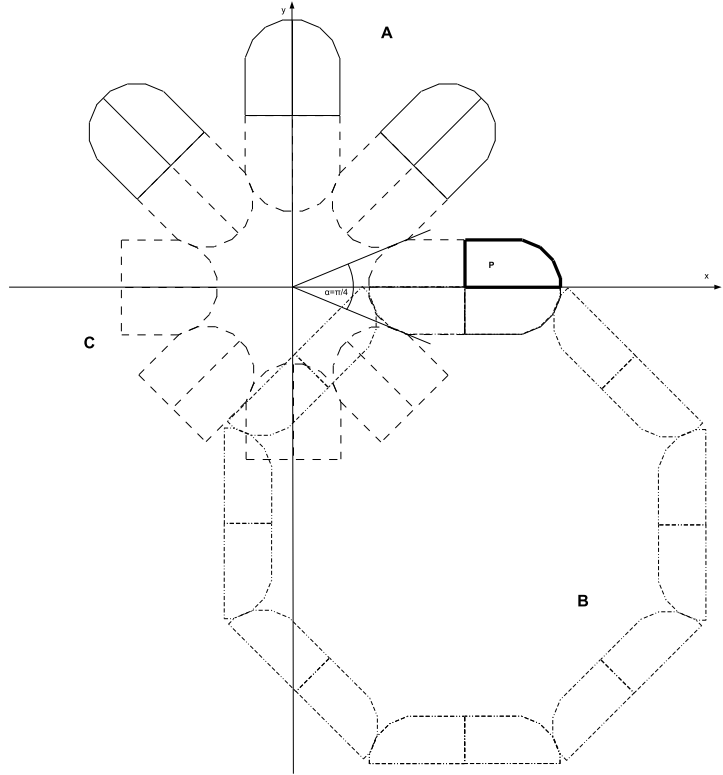,width=10 cm} \caption{Some three equivalent forms {\bf A}, {\bf B} and {\bf C} of EPPs for the polygon billiards $A''$. The form
{\bf B} as well as {\bf C} coincide partly with the {\bf A} one. The latter form is used in our further considerations}
\end{center}
\end{figure}

\subsection{Semiclassical wave functions built in the polygon billiards $A''$}

\hskip+2em According to the general rules \cite{53} governing the construction of the semiclassical wave functions (SWF) for the case considered we have
to define
first an elementary polygon pattern (EPP) on which such a SWF is built. There are many possible EPPs for a given RPB three examples {\bf A}, {\bf B} and
{\bf C} of which are shown in Fig.2. Below the EPP {\bf A} will be used in our further considerations, see Fig.3A. It was built by the subsequent mirror
reflections of the polygon $P$ in its sides $a$, $b$ and $c$, see Fig.3B. The original orientation of the polygon $A''$ is denoted in Fig.3A by the
letter $P$ and the sign $"+"$ while its "odd" mirror reflections by $"-"$. The basic property of any EPP is that any additional mirror reflection of
the polygon with any orientation (of its two possible) in any of the side of the EPP always recovers some of the polygons it contains, i.e. the EPP is
the maximal construction made of the reflected polygons in which their all positions and orientations are not repeatable.

\begin{figure}
\begin{center}
\psfig{figure=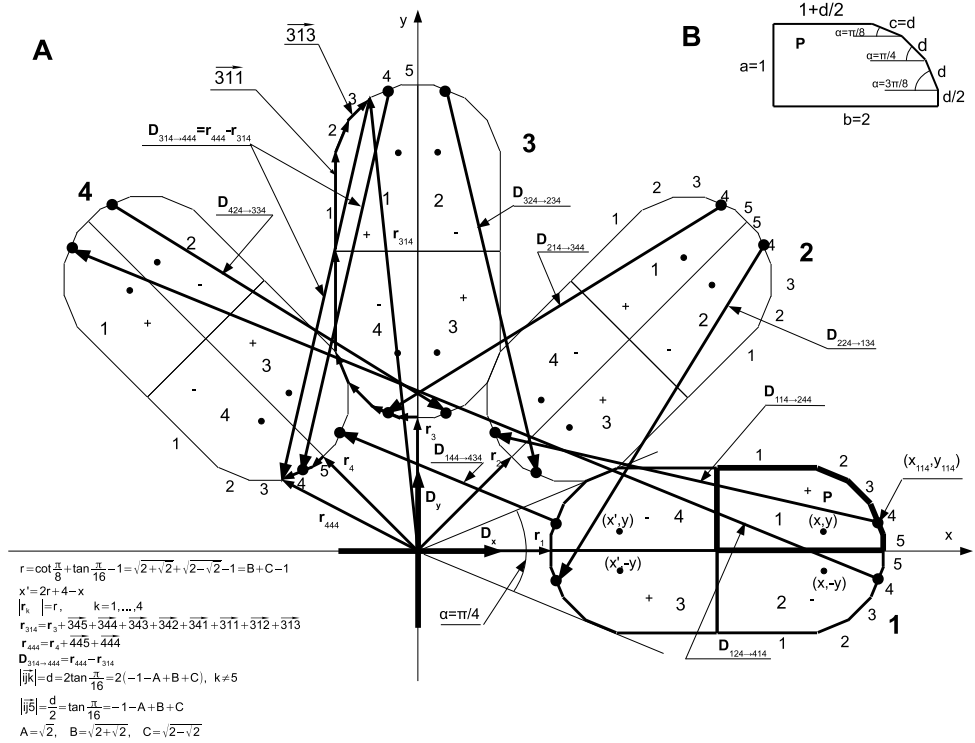,width=14 cm} \caption{The EPP for the polygon billiards $A''$. The eight periods shown in the figure link the boundary point
$(x^{114},y^{114})$ and its seven images with the remaining eight ones. All the image points contribute to the SWF at the point $(x^{114},y^{114})$ for
both the polygon billiards $P$ and the Bunimovich one}
\end{center}
\end{figure}

Continuing however mirror reflections of the polygon billiards $P$ outside its fixed EPP by any of its sides we get a complicated surface made of all such
mirror reflections called the rational polygon Riemann surface (RPRS) \cite{41}. The latter can always be recover by translations of the EPP chosen by
all possible periods of the RPRS. Nevertheless the structure of RPRS is independent of the chosen EPP used to its construction mentioned.

In the next step we should define a full number of independent periods corresponding to the EPP of Fig.3A. Since the polygon considered is
pseudointegrable with the corresponding genus $g=13$ of the respective multi-torus the number of independent periods mentioned is equal to $26$.
There are shown $8$ of them on Fig.3A. The periods can be identified by applying
to the EPP the basic rule by which each period links a pair of two parallel sides of EPP belonging to a pair of two different polygons having opposite
orientations. Each of the two such polygons can be get from the other by its mirror reflection in the side belonging to the pair mentioned and translating the
image just by the period wanted. On Fig.3A there is also shown the way by which the respective periods are enumerated.

The subsequent step is to choose two periods independent on the $x,y$-plane and to express the remaining ones as approximate linear combinations of them
with respective rational coefficients approximating the proper real ones. Choosing the periods ${\bf D}_x={\bf D}_{311\to 321}$ and
${\bf D}_y={\bf D}_{121\to 111}$ we have for the exact relations
\be
{\bf D}_{ilk\to rst}=a_x^{ilk\to rst}{\bf D}_x+a_y^{ilk\to rst}{\bf D}_y\nn\\
\label{1}
\ee
between all $29$ relevant periods of the EPP of Fig.3A and the periods ${\bf D}_x,\;{\bf D}_y$.

The detailed forms of the real coefficients $a_x^{ijk\to rst},a_y^{ijk\to rst}$ can be obtained by studying geometrical relations between periods determined by the EPP of
Fig.3A. Doing this (see App.A.1) it can however be observed that each of these coefficients can be linearly expressed by the four real numbers
$1,\;\sqrt{2},\;\sqrt{2-\sqrt{2}},\;\sqrt{2+\sqrt{2}}$ with rational coefficients and the denominators of these rationals have as their least common
multiple the number 4. Therefore to approximate the coefficients $a_x^{ijk\to rst},a_y^{ijk\to rst}$ by rationals it is enough to approximate by rationals the last three
irrationals mentioned with a controlled and desired level of an accuracy. Putting therefore
$X_0=1,\;X_1=\sqrt{2},\;X_2=\sqrt{2+\sqrt{2}},\;X_3=\sqrt{2-\sqrt{2}}$ we can write
\be
a_x^{ilk\to rst}=\frac{1}{4}\sum_{f=0}^3a_{xf}^{ilk\to rst}X_f,\;\;a_y^{ilk\to rst}=\frac{1}{4}\sum_{f=0}^3a_{yf}^{ilk\to rst}X_f
\label{3e}
\ee
where $a_{xi}^{ijk\to rst},\;a_{yi}^{ijk\to rst},\;i=1,2,3,$ are integer.

Next using the Dirichlet simultaneous approximation theorem (see App.B) we have
\be
|ZX_k-q_k|<\frac{1}{N^\frac{1}{3}}\nn\\
0<Z<N,\;\;\;\;k=1,2,3
\label{2a}
\ee
where $q_k,\;k=1,2,3,\;Z,\;N$ are all natural and $N$ is arbitrary.

Having the multiplier $Z$ we can write next the following quantization conditions for the momentum ${\bf p}=[p_x,p_y]$ of the billiards
ball
\be
{\bf p}\cdot{\bf D}_x=p_xD_x=8\pi mZ\nn\\
{\bf p}\cdot{\bf D}_y=p_yD_y=8\pi nZ\nn\\
m,n=0,\pm 1,\pm 2,...,\;\;\;\;|m|+|n|>0
\label{3}
\ee
with $D_x=D_y=2$ being the lengths of each of the two periods ${\bf D}_x$ and ${\bf D}_y$ independent on the plane and corresponding to the bouncing
ball periodic motion between the flat sides of the Bunimovich stadium.

Then for each period ${\bf D}_{ijk\to rst}$, we have
\be
{\bf p}\cdot{\bf D}_{ilk\to rst}=8\pi\ll(mZa_x^{ilk\to rst}+nZa_y^{ilk\to rst}\r)=2\pi\sum_{f=0}^3(ma_{xf}^{ilk\to rst}+na_{yf}^{ilk\to rst})ZX_f
\label{3d}
\ee
so that
\be
\ll|{\bf p}\cdot{\bf D}_{ilk\to rst}-2\pi I_{mn}^{ilk\to rst}\r|<2\pi\frac{|m|I_x^{ilk\to rst}+|n|I_y^{ilk\to rst}}{N^\frac{1}{3}}\nn\\
I_{mn}^{ilk\to rst}=m\sum_{f=0}^3a_{xf}^{ilk\to rst}q_f+n\sum_{f=0}^3a_{yf}^{ilk\to rst}q_f\nn\\
I_x^{ilk\to rst}=\sum_{f=1}^3\ll|a_{xf}^{ilk\to rst}\r|,\;\;I_y^{ilk\to rst}=\sum_{i=f}^3\ll|a_{yf}^{ilk\to rst}\r|\nn\\
m,n=0,\pm 1,\pm 2,...,
\label{3a}
\ee
where $I_{mn}^{ijk\to rst}$ is integer while $I_x^{ijk\to rst},\;I_y^{ijk\to rst}$ are positive integers.

The respective energy spectrum provided by \mref{3} is
\be
E_{mn}=\fr{\bf p}^2=8\pi^2Z^2(m^2+n^2)
\label{3b}
\ee

It is important to note at this moment that the main conclusions done in the above procedure would remain unchanged if one chose another pair of two independent
periods instead of ${\bf D}_x$ and ${\bf D}_y$. A possible change would touch only the coefficients of the linear relations \mref{3e} and values of
their least common multiples while the four linear independent irrationals $X_i$ would stay unchanged, i.e. the number $Z$ in \mref{2a} would be the
same but the coefficient $8\pi$ in the quantization conditions \mref{3} could be changed to $2\pi\times w$ where $w$ would be a new least common multiple
of denominators of the coefficients in \mref{3e}.

The insensitivity of the irrationals $X_i$ on the base periods changes is ensured by their algebra (see App.A.4) but possible changes of least common multiples
mean however that the energy spectra given by \mref{3b} can also be changed, i.e. different choices of the base periods can provide us with
different domains of approximated energy spectra of the polygon billiards and therefore also with different SWFs accompanied them.

The next step in the routine procedure of constructing SWFs corresponding to the EPP of Fig.2A is to take so called basic semiclassical wave function
(BSWF) \cite{41} of the form
\be
\Psi^{BSWF}(x,y)=\pm e^{i(p_xx+p_yy)}
\label{4}
\ee
and to sum it over all images of the point $(x,y)$ of the billiards got by forming the EPP of Fig.2A and attaching to it the corresponding signs shown
in the figure.

The arrangement of the signs shown in Fig.2A corresponds to the construction of approximate SWFs in the polygon billiards $A''$ satisfying
the Dirichlet conditions on the boundary of the billiards which farther corresponds to the respective SWFs in the Bunimovich stadium which are
antisymmetric with respect to its both symmetry axes. The choice of the Dirichlet conditions is however not arbitrary since a system of signs which
could provide us with other boundary conditions guaranteeing symmetry properties of SWFs in the Bunimovich stadium other than the antisymmetric ones
is not available in our approach.

Therefore in this way we get
\begin{enumerate}
\item an approximate semiclassical solution corresponding to the Dirichlet boundary conditions both for the polygon
billiards $A''$ of Fig.1 and for the polygon billiards $A'$ of the figure composed of the former four ones; and
\item SWFs which are strictly antisymmetric with respect to both the symmetry axes of the polygon billiards $A'$
\end{enumerate}

The latter property the polygon billiards $A'$ gets due to the choice of the EPP $A$ of Fig.2, i.e. a choice of other EPPs of the figure would not
lead us directly to SWFs for the billiards $A'$ with their second property mentioned above. This is because the quantized polygon billiards $A''$
is not of the doubly rational polygon billiards (DRPB) class for which all linear relations on the plane between their independent periods have only
rational coefficients \cite{42}. If it was then any choice of allowed EPP for the billiards would be irrelevant for its semiclassical quantization
contrary to the considered case which being not DRPB one makes differences in the resulting quantizations because of their dependence on the
approximations \mref{2a} of irrationals by respective rationals.

Using therefore \mref{4} we get (up to a normalization constant)
\be
\Psi_{mn}^{sem}(x,y)=-\frac{1}{4}\sum_{over EPP}\pm e^{i(p_xx_i+p_yy_i)}=\nn\\
                     e^{4\pi mZ(r+2)}\sin(4m\pi Z(x-r-2))\sin(4n\pi Zy)-\nn\\
                     e^{4\pi nZ(r+2)}\sin(4n\pi Z(x-r-2))\sin(4m\pi Zy)+\nn\\
                     e^{2\pi i(m-n)\sqrt{2}Z(r+2)}\sin\ll(2(m-n)\pi\sqrt{2}Z(x-r-2)\r)\sin\ll(2(m+n)\pi\sqrt{2}Zy\r)-\nn\\
                     e^{2\pi i(m+n)\sqrt{2}Z(r+2)}\sin\ll(2(m+n)\pi\sqrt{2}Z(x-r-2)\r)\sin\ll(2(m-n)\pi\sqrt{2}Zy\r)
\label{5}
\ee

Obviously, because of \mref{3a} both the BSWF \mref{4} and the SWF \mref{5} are not periodic with respect to the periods
${\bf D}_{ijk\to rst}$. As a consequence of this while $\Psi^{sem}(x,y)$ vanishes by its construction on the two sides $a$ and $b$ of the
polygon billiards $P$ used to built its EPP it does not vanish on the remaining ones. This is just the consequence of that the periodic
trajectories of the polygon considered are not in general integer multiples of the lengths of waves corresponding to their periods (note by the way that a
periodic trajectory reflecting $r$ times off the billiards
boundary generates $r$ periods of the same length each). In fact if ${\bf p}\cdot{\bf D}_{ijk\to rst}=\pm p_{ijk\to rst}D_{ijk\to rst}=
\pm 2\pi D_{ijk\to rst}/\lambda_{ijk\to rst}$ we can rewrite \mref{3a} as
\be
\ll|D_{ilk\to rst}-|I_{mn}^{{ilk\to rst}}|\lambda_{ilk\to rst}\r|<
\frac{|m|I_x^{{ilk\to rst}}+|n|I_y^{{ilk\to rst}}}{N^\frac{1}{3}}\lambda_{ilk\to rst}
\label{6}
\ee

It is therefore clear that an application of the above way of the semiclassical quantization to the considered case of the RPB makes sens only when
the rho of \mref{6} is sufficiently close to zero for every period since only then the SWF \mref{5} can be close to zero on the boundary of
the RPB of Fig.1$A''$ to satisfy at least approximately the Dirichlet boundary conditions as it can be seen from the following calculations.

First let us note that in the chosen EPP on Fig.3A the mirror reflections of the polygon $P$ by its sides $a$ and $b$ form four other polygons each of which is exactly the
polygon envelope $A'$ of the Bunimovich stadium $A$ of Fig.1. Therefore the SWFs \mref{5} are immediately such functions for the polygon billiards
$A'$ too. Note however that the EPP of Fig.3A is not an EPP for the billiards $A'$.

Let us now denote by $(x_{ijk},y_{ijk})$ a point lying on the side $ijk$ of the RPB $P$ boundary according to the respective enumeration of the periods
${\bf D}_{ijk\to rst}$, emerging from this side as it is shown in Fig.3A. There are still three other points of the polygon {\bf 1} and twelve farther points
of the EPP contributing to SWFs at the point ${\bf r}_{ijk}=(x_{ijk},y_{ijk})$. But only half of them contribute independently of the others the latter being related
with the previous ones by respective periods.

Let us demonstrate the respective contributions to SWFs \mref{5} at the point $(x_{114},y_{114})$ of the
polygon $P$ boundary. We have
\be
\ll|\Psi_{mn}^{sem}(x_{114},y_{114})\r|=\nn\\
\ll|e^{i{\bf p}{\bf r}_{114}}-e^{i{\bf p}({\bf r}_{114}+{\bf D}_{114\to 244})}-
e^{i{\bf p}{\bf r}_{124}}+e^{i{\bf p}({\bf r}_{124}+{\bf D}_{124\to 414})}-
    e^{i{\bf p}{\bf r}_{224}}+e^{i{\bf p}({\bf r}_{224}+{\bf D}_{224\to 134})}\r.-\nn\\
\ll.    e^{i{\bf p}{\bf r}_{144}}+e^{i{\bf p}({\bf r}_{144}+{\bf D}_{144\to 434})}+
    e^{i{\bf p}{\bf r}_{214}}-e^{i{\bf p}({\bf r}_{214}+{\bf D}_{214\to 344})}+
    e^{i{\bf p}{\bf r}_{324}}-e^{i{\bf p}({\bf r}_{324}+{\bf D}_{324\to 234})}+\r.\nn\\
\ll.    e^{i{\bf p}{\bf r}_{314}}-e^{i{\bf p}({\bf r}_{314}+{\bf D}_{314\to 444})}-
    e^{i{\bf p}{\bf r}_{424}}+e^{i{\bf p}({\bf r}_{424}+{\bf D}_{424\to 334})}\r|\leq\nn\\
2\ll(\ll|\sin\ll(\fr({\bf p}{\bf D}_{114\to 244}-2\pi I_{mn}^{114\to 244}\r)\r|+\ll|\sin\ll(\fr({\bf p}{\bf D}_{124\to 414}-2\pi I_{mn}^{124\to 414}\r)\r|\r.+\nn\\
\ll.\ll|\sin\ll(\fr({\bf p}{\bf D}_{224\to 134}-2\pi I_{mn}^{224\to 134}\r)\r|+\ll|\sin\ll(\fr({\bf p}{\bf D}_{144\to 434}-2\pi I_{mn}^{144\to 434}\r)\r|\r.+\nn\\
\ll.\ll|\sin\ll(\fr({\bf p}{\bf D}_{214\to 344}-2\pi I_{mn}^{214\to 344}\r)\r|+\ll|\sin\ll(\fr({\bf p}{\bf D}_{324\to 234}-2\pi I_{mn}^{324\to 234}\r)\r|\r.+\nn\\
\ll.\ll|\sin\ll(\fr({\bf p}{\bf D}_{314\to 444}-2\pi I_{mn}^{314\to 444}\r)\r|+\ll|\sin\ll(\fr({\bf p}{\bf D}_{424\to 334}-2\pi I_{mn}^{424\to 334}\r)\r|\r)<\nn\\
  2\pi\frac{|m|J_x^{114}+|n|J_y^{114}}{N^{\frac{1}{3}}}\nn\\
    J_k^{114}=I_k^{114\to 244}+I_k^{124\to 414}+I_k^{224\to 134}+ I_k^{144\to 434}+I_k^{214\to 344}+ I_k^{324\to 234}+\nn\\
    I_k^{314\to 444}+I_k^{424\to 334},\;\;\;\;\;\;    k=x,y
\label{7}
\ee
where we have taken into account \mref{3a}.

Since $N$ can be taken arbitrarily large then $\ll|\Psi_{mn}^{sem}(x_{ijk},y_{ijk})\r|$ can be done arbitrarily small on each side of the billiards
$A''$ and $A'$ of Fig.1 for a set $\{m,n:|m|J_x^{ijk}+|n|J_y^{ijk}<<N^{\frac{1}{3}}\}$.

Nevertheless to get in the relation \mref{7} an accuracy say $10^{-4}$ we have still to inspect $N=10^{12}$ first naturals looking for
the number $Z$ among them. Table \ref{table:A} below shows a dependence of $N$ and $Z$ on the respective accuracies. The corresponding results have
been computed with the double precisions by Fortran 95 for PC which allows us for maximal $N=2\times 10^9$ for which the number we get the third row of the
table. However one can still look for the best accuracy given by the last row of the table. The accuracies better than $3,60\times 10^{-4}$
cannot be achieved by computing with Fortran 95 for PC, i.e. the respective $Z$s have not been found within the range of integers provided by Fortran
95. The accuracies in Table \ref{table:A} corresponds to the smallest $Z$ given in the table, i.e. decreasing by one the last decimal place of an
accuracy from a row of the table increases $Z$ to its value in the next row. It means that $Z$ is a decreasing step function of accuracy as it is shown
in Fig.4.

It is clear that each computed $Z$ can be used in the SWF \mref{5} defining the SWF itself and a region of energy spectrum approximated by the formula
\mref{3b} with the respective accuracy given by Table \ref{table:A}. However the better the approximation is to be the larger $Z$ has to be used and the
higher regions of the energy spectrum is then approximated by \mref{3b}.

\begin{table}[b]
\caption{A dependence of $N$ and $Z$ on accuracy in the case $A$}
\centering
\begin{tabular}{r|r|r|r|r|r}
\hline\hline
accuracy&N=(accuracy)$^{-3}$&Z&$q_1$&$q_2$&$q_3$\\
\hline
$8,67\times 10^{-4}$&68702736&186445124&263673223&344505668&142698920\\
$8,62\times 10^{-4}$&69905207&287348498&406372143&530950792&219927019\\
$7.94\times 10^{-4}$&2000000000&937322935&1325574807&1731946950&717395916\\
$6,40\times 10^{-4}$&3814697266&937322935&1325574807&1731946950&717395916\\
$3,60\times 10^{-4}$&21433470508&1038226309&1468273727&1918392074&794624015
\end{tabular}
\label{table:A}
\end{table}

\begin{figure}
\begin{center}
\psfig{figure=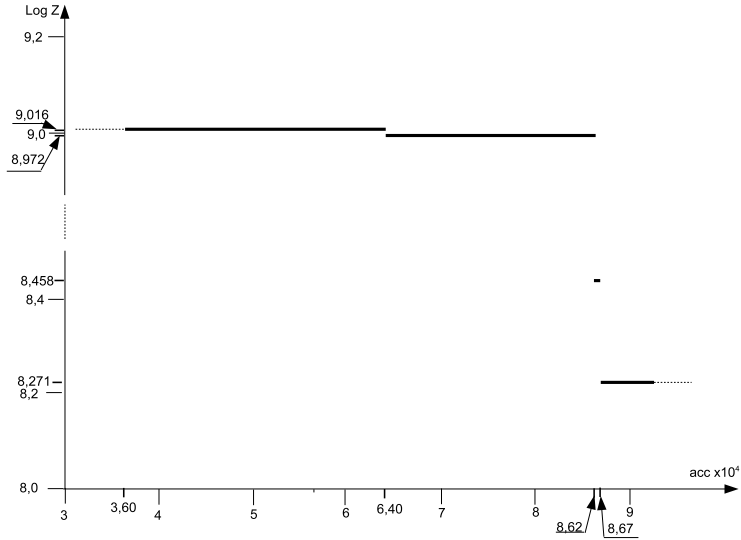,width=10 cm} \caption{The step function dependence of $Z$ on the accuracy for the polygon billiards $A''$}
\end{center}
\end{figure}

The above discussion shows also that the complexness of the SWF suggesting by \mref{5} is to some extent apparent. Namely it is seen from the last
estimations that only the real part of \mref{5} can be relevant in sufficiently high energy regions since then the imaginary one is close to zero, i.e.
every of the four coefficients in \mref{5} is then close to unity.

The same arguments allow us to consider the energy levels $E_{mn}$ as not being degenerate as it would be suggested by the formula \mref{3b} which is
insensitive on signs of the quantum numbers $m,n$. However the differences between the SWFs $\Psi_{mn}^{sem}(x,y)$ given by \mref{5} and corresponding
to the quantum numbers $m,n$ differing only by their signs are also limited by numbers of the order $N^{-1/3}$, i.e. up to such an accuracy we can
consider all $\Psi_{\pm|m|\pm|n|}^{sem}(x,y)$ as equal up to a sign.

\subsection{Semiclassical wave functions built in the polygon billiards $B''$ and $C''$}

\hskip+2em We consider both the cases together because of their close similarity to each other (see Fig.Fig.5,6), i.e. possible differences between them
are reduced in fact to different values of some basic parameters describing the cases. We can follow very closely to the procedure of the previous subsection
taking into account that the main difference appears as a larger
number of independent periods defined by the present cases. The genus $g$ of the multi-torus which corresponds to the case considered is equal to 33 so
that 66 is the total number of independent periods. One can identify them among 73 periods which link pairs of the parallel sides of the EPPs of Fig.5
and 6. The latter periods can be represented on the $x,y$-plane as the linear combinations of the periods ${\bf D}_x$ and ${\bf D}_y$ by

\begin{figure}
\begin{center}
\psfig{figure=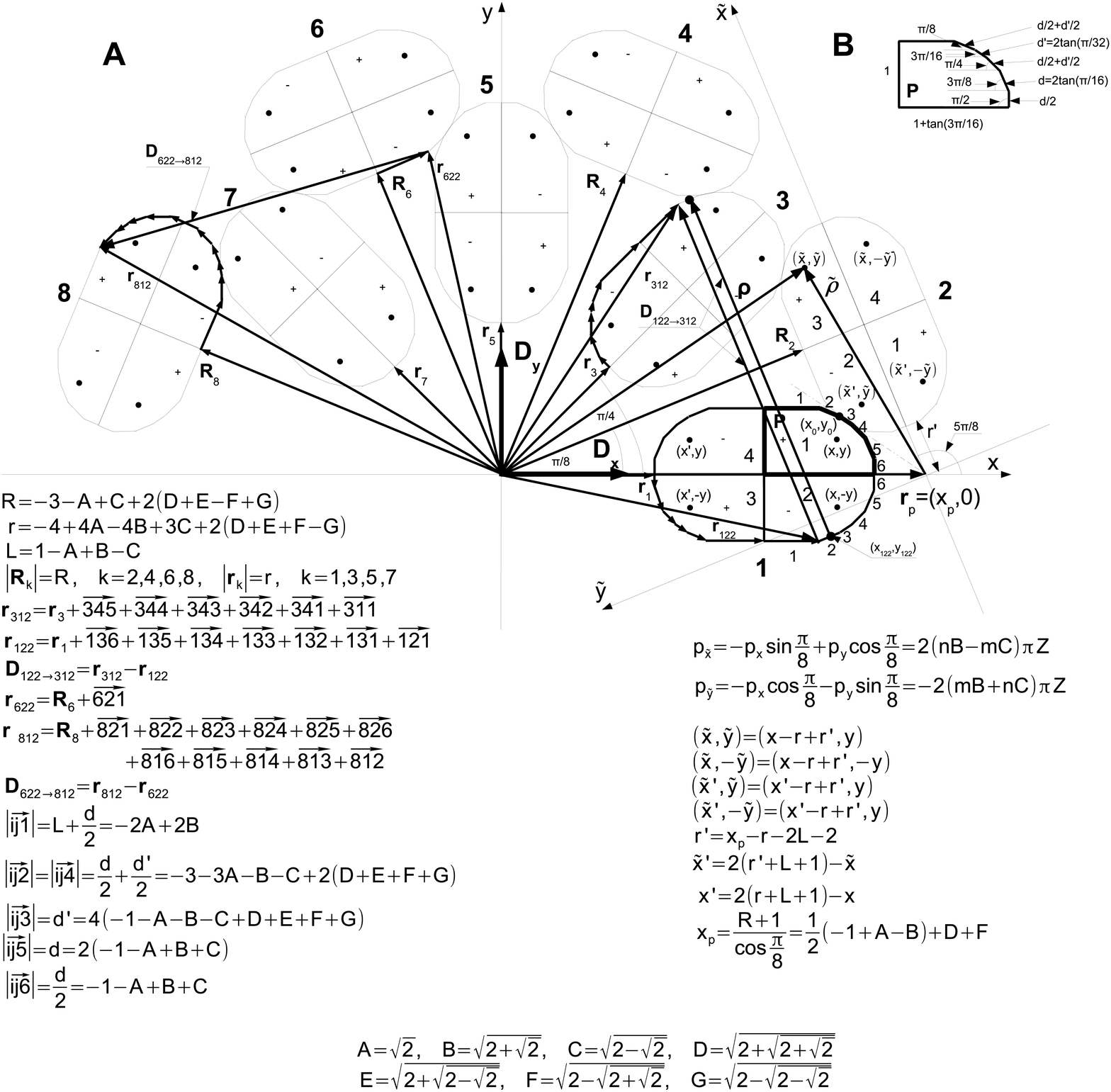,width=10 cm} \caption{EPP for the polygon billiards $B''$}
\end{center}
\end{figure}

\begin{figure}
\begin{center}
\psfig{figure=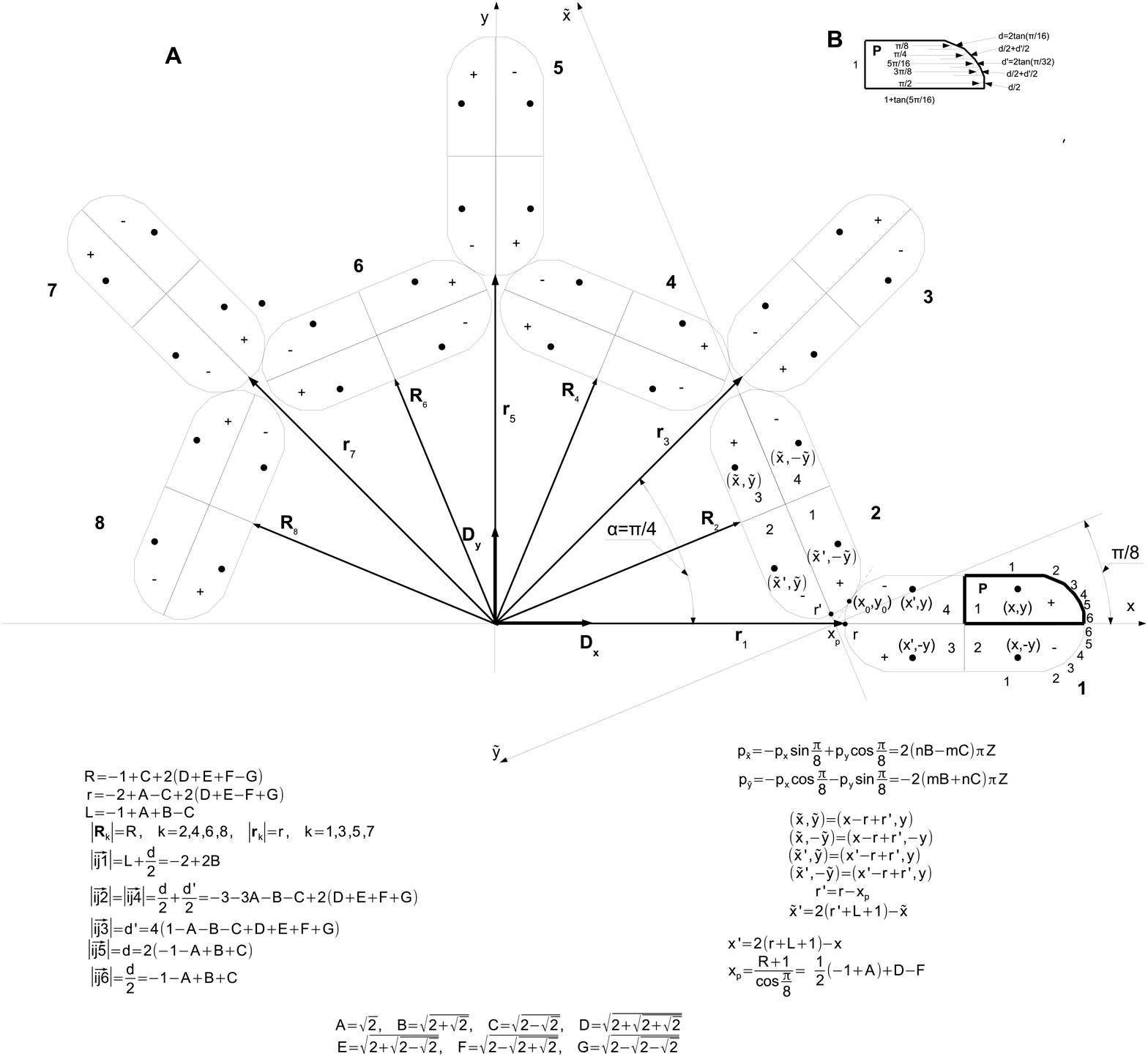,width=10 cm} \caption{EPP for the polygon billiards $C''$}
\end{center}
\end{figure}

\be
{\bf D}_{ilk\to rst}=a_x^{ilk\to rst}{\bf D}_x+a_y^{ilk\to rst}{\bf D}_y\nn\\
a_k^{ilk\to rst}=\frac{1}{4}\sum_{q=0}^7a_{kq}^{ilk\to rst}X_q,\;\;\;k=x,y\nn\\
X_0=1,\;X_1=A,\;X_2=B,\;X_3=C,\;
X_4=D=\sqrt{2+\sqrt{2+\sqrt{2}}},\nn\\
X_5=E=\sqrt{2+\sqrt{2-\sqrt{2}}},\;
X_6=F=\sqrt{2-\sqrt{2+\sqrt{2}}},\;X_7=G=\sqrt{2-\sqrt{2-\sqrt{2}}}
\label{A2}
\ee
and according to App.A.2 $a_{kq}^{ijk\to rst}$ are all integer.

Farther according to the Dirichlet theorem (see App.B) the seven independent irrationals $X_i,\;i=1,...,7$, which govern the linear
relations between the periods (some of them are shown in Fig.4) can be approximated simultaneously by rationals as follows
\be
|ZX_i-q_i|<\frac{1}{N^\frac{1}{7}},\;\;\;\;\;\;\;\;i=1,...,7
\label{8}
\ee
for an arbitrary natural $N$, natural $Z,\;Z<N$, and integer $q_i,\;i=1,...,7$.

In fact, for $N=2\times 10^9$ one gets $N^\frac{1}{7}=4,691\times 10^{-2}$ so that such $N$ does not guarantee too good approximation. However
making computations similar to the ones from the previous sections we get results summarized in the Tables \ref{table:B} and \ref{table:C} below
and in Fig.7.

\begin{table}[b]
\caption{Dependence of $N$ and $Z$ on accuracy for the cases $B$ and $C$}
\centering
\begin{tabular}{|r|r|r|r|r}
\hline\hline
accuracy&N=(accuracy)$^{-7}$&Z&$q_1$&$q_2$\\
\hline
$4,951\times 10^{-2}$&1371353804&6743502&9536752&12460367\\
$4,691\times 10^{-2}$&2000000000&8019788&11341693&14818636\\
$4,685\times 10^{-2}$&2018518962&8019788&11341693&14818636\\
$4,356\times 10^{-2}$&3360353005&11214034&15859039&20720833\\
$3,316\times 10^{-2}$&22682887919&31934867&45162722&59007940\\
$2,905\times 10^{-2}$&57276575342&226662402&320549043&418817508\\
$2,825\times 10^{-2}$&69641935871&287337890&406357141&530931191
\end{tabular}
\label{table:B}
\end{table}

\begin{table}[t]
\caption{Table \ref{table:B} continued}
\centering
\begin{tabular}{r|r|r|r|r|}
\hline\hline
$q_3$&$q_4$&$q_5$&$q_6$&$q_7$\\
\hline
5161253&13227855&11214034&2631184&7492978\\
6138080&15731380&13336420&3129166&8911111\\
8582850&21997119&18648257&4375499&12460367\\
24441889&62642495&53105743&12460367&35484123\\
173479892&444614295&376925799&88439282&251853767\\
219918900&563633546&477825448&112113683&319272757
\end{tabular}
\label{table:C}
\end{table}

\begin{figure}
\begin{center}
\psfig{figure=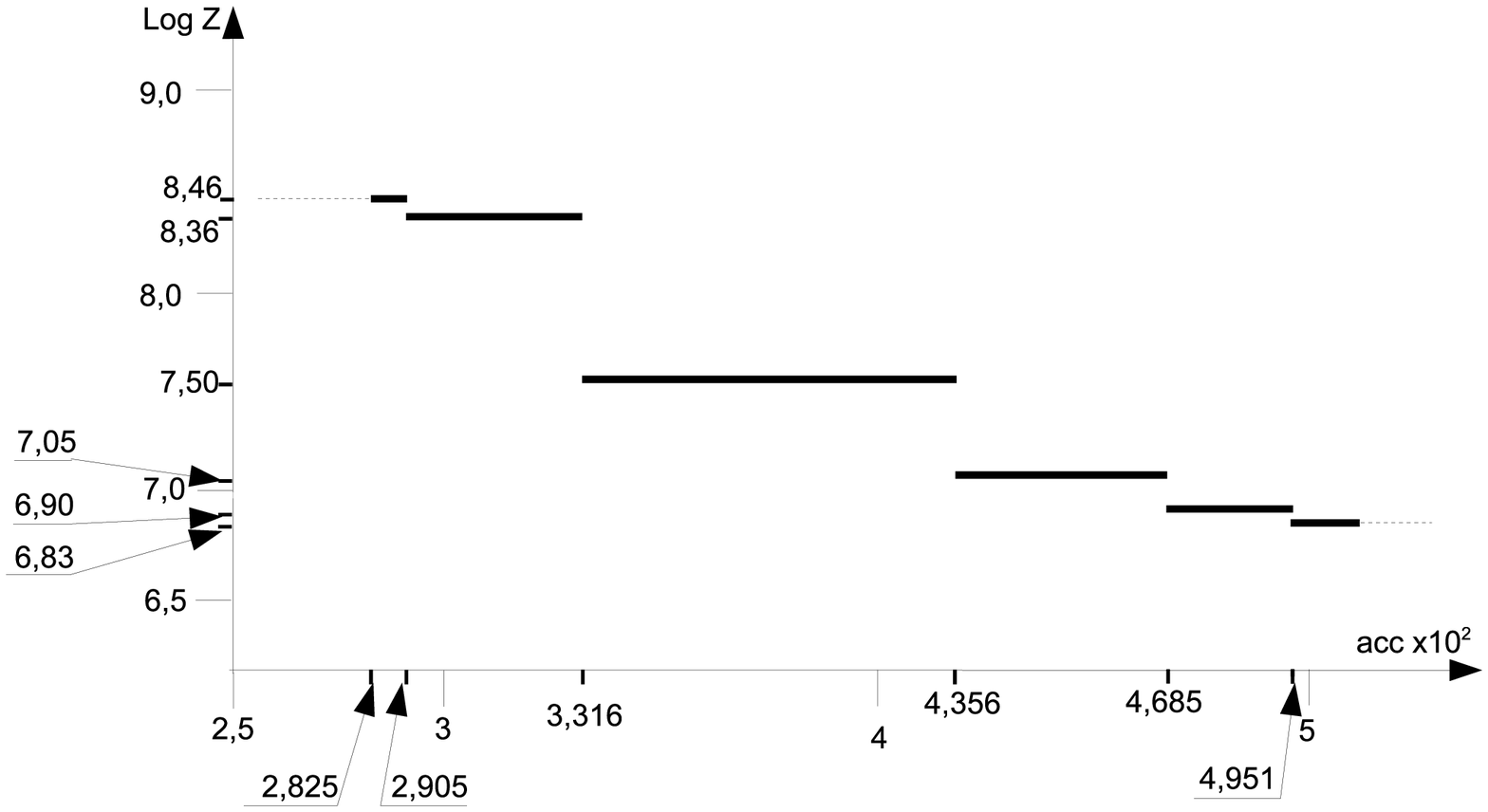,width=13 cm} \caption{The step function dependence of $Z$ on the accuracy for the polygon billiards $B''$ and $C''$}
\end{center}
\end{figure}

Since the rational coefficients in the linear relations \mref{A2} have the number 4 as their least common multiple we can write the following
approximate quantization conditions for the momenta in the billiards considered
\be
{\bf p}\cdot{\bf D}_x=8\pi mZ\nn\\
{\bf p}\cdot{\bf D}_y=8\pi nZ\nn\\
m,n=0,\pm 1,\pm 2,...
\label{9}
\ee
while for any period \[{\bf D}_{ijk\to rst}=a_x^{ijk\to rst}{\bf D}_x+a_y^{ijk\to rst}{\bf D}_y=\frac{1}{4}\sum_{f=0}^7(a_{xf}^{ijk\to rst}{\bf D}_x+
a_{yf}^{ijk\to rst}{\bf D}_y)X_f\] of Fig.Fig.4,5 the respective conditions is satisfied as
\be
{\bf p}\cdot{\bf D}_{ilk\to rst}=\frac{1}{4}\sum_{f=0}^7(a_{xf}^{ilk\to rst}{\bf p}\cdot{\bf D}_x+a_{yf}^{ilk\to rst}{\bf p}\cdot{\bf D}_y)Z_f=\nn\\
2\pi\sum_{f=0}^7(ma_{xf}^{ilk\to rst}+na_{yf}^{ilk\to rst})ZX_f
\label{10}
\ee
so that
\be
|{\bf p}\cdot{\bf D}_{ilk\to rst}-2\pi I_{mn}^{ilk\to rst}|<2\pi(|m|I_x^{ilk\to rst}+|n|I_y^{ilk\to rst})\frac{1}{N^\frac{1}{7}}\nn\\
I_{mn}^{ilk\to rst}=m\sum_{f=0}^7a_{xf}^{ilk\to rst}q_f+n\sum_{f=0}^7a_{yf}^{ilk\to rst}q_f\nn\\
I_x^{ilk\to rst}=\sum_{f=1}^7|a_{xf}^{ilk\to rst}|,\;
I_y^{ilk\to rst}=\sum_{f=1}^7|a_{yf}^{ilk\to rst}|
\label{11}
\ee
where $I_x^{ijk\to rst},I_y^{ijk\to rst}$ are positive integers while $I_{mn}^{ijk\to rst}$ are integer.

Therefore the semiclassical quantization of energy levels gives us
\be
E_{mn}=8\pi^2Z^2(m^2+n^2)\nn\\
m,n=0,\pm 1,\pm 2,...
\label{12}
\ee
and the corresponding SWFs are following (up to a normalization constant)
\be
\Psi_{mn}^{sem}(x,y)=-\frac{1}{4}\sum_{over EPP}\pm e^{i(p_xx_i+p_yy_i)}=\nonumber\\
                     e^{4\pi mZ(r+L+1)}\sin(4\pi mZ(x-r-L-1))\sin(4\pi nZy)-\nonumber\\
                     e^{2\pi i(nB-m(C-2x_p))Z(r'+L+1)}\times\nonumber\\
                     \sin(2\pi(nB-mC)Z(x-r-L-1))\sin(2\pi(mB+nC)Zy)-\nonumber\\
                     e^{4\pi nZ(r+L+1)}\sin(4\pi nZ(x-r-L-1))\sin(4\pi mZy)+\nonumber\\
                     e^{-2\pi i(mB+n(C-2x_p))Z(r'+L+1)}\times\nonumber\\
                     \sin(2\pi(mB+nC)Z(x-r-L-1))\sin(2\pi(nB-mC)Zy)+\nonumber\\
                     e^{-2\pi i(m-n)\sqrt{2}Z(r+L+1)}\sin(2\pi(m-n)\sqrt{2}Z(x-r-L-1))\sin(2\pi(m+n)\sqrt{2}Zy)-\nonumber\\
                     e^{-2\pi i(n(B-x_p)+m(C+x_p))Z(r'+L+1)}\times\nonumber\\
                     \sin(2\pi(nB+mC)Z(x-r-L-1))\sin(2\pi(mB-nC)Zy)+\nonumber\\
                     e^{2\pi i(m+n)\sqrt{2}Z(r+L+1)}\sin(2\pi(m+n)\sqrt{2}Z(x-r-L-1))\sin(2\pi(m-n)\sqrt{2}Zy)+\nonumber\\
                     e^{-2\pi i(m(B-x_p)-n(C+x_p))Z(r'+L+1)}\times\nonumber\\
                     \sin(2\pi(mB-nC)Z(x-r-L-1))\sin(2\pi(nB+mC)Zy)\;\;\;\;\;
\label{13}
\ee
where the parameters $L,r,r',x_p$ are given on the respective figures 5 and 6.

The estimations of accuracies of $\Psi_{mn}^{sem}(x,y)$ by checking its closeness to zero on the boundaries of the billiard
considered can be done along the same lines as in the previous case (see the formula \mref{7}) so that taking into account the estimations \mref{8}
we can write for any boundary point $(x_{11k},y_{11k})$ of the polygons $B',C'$
\be
|\Psi_{mn}^{sem}(x_{11k},y_{11k})|<2\pi(|m|J_x^{11k}+|n|J_y^{11k})\frac{1}{N^\frac{1}{7}},\;\;\;\;\;\;k=1,...,6
\label{14}
\ee
where $J_x^{11k},J_y^{11k}$, are defined by formulas similar to \mref{7} taking into account all possible images of the point considered and the
respective periods ${\bf D}_{ijk\to rst}$ linking them.

The comments finishing the previous subsection can be also repeated here with no changes.

\section{SWFs built on a periodic skeleton - superscars phenomena in the polygon billiards enveloping the Bunimovich stadium}

\hskip+2em In the previous section the SWFs \mref{5} and \mref{13} were built on aperiodic skeletons giving us the general form of the
quantization conditions \mref{3}, \mref{12}
as well as of the SWFs themselves. However the quantization procedure can be performed equally well on periodic skeletons if there are no constraints which
can prevent such a quantization \cite{42}. Fortunately in the cases of the considered RPBs the respective constraints can be easily satisfied without
any condition on the properties of the polygons.

\begin{figure}
\begin{center}
\psfig{figure=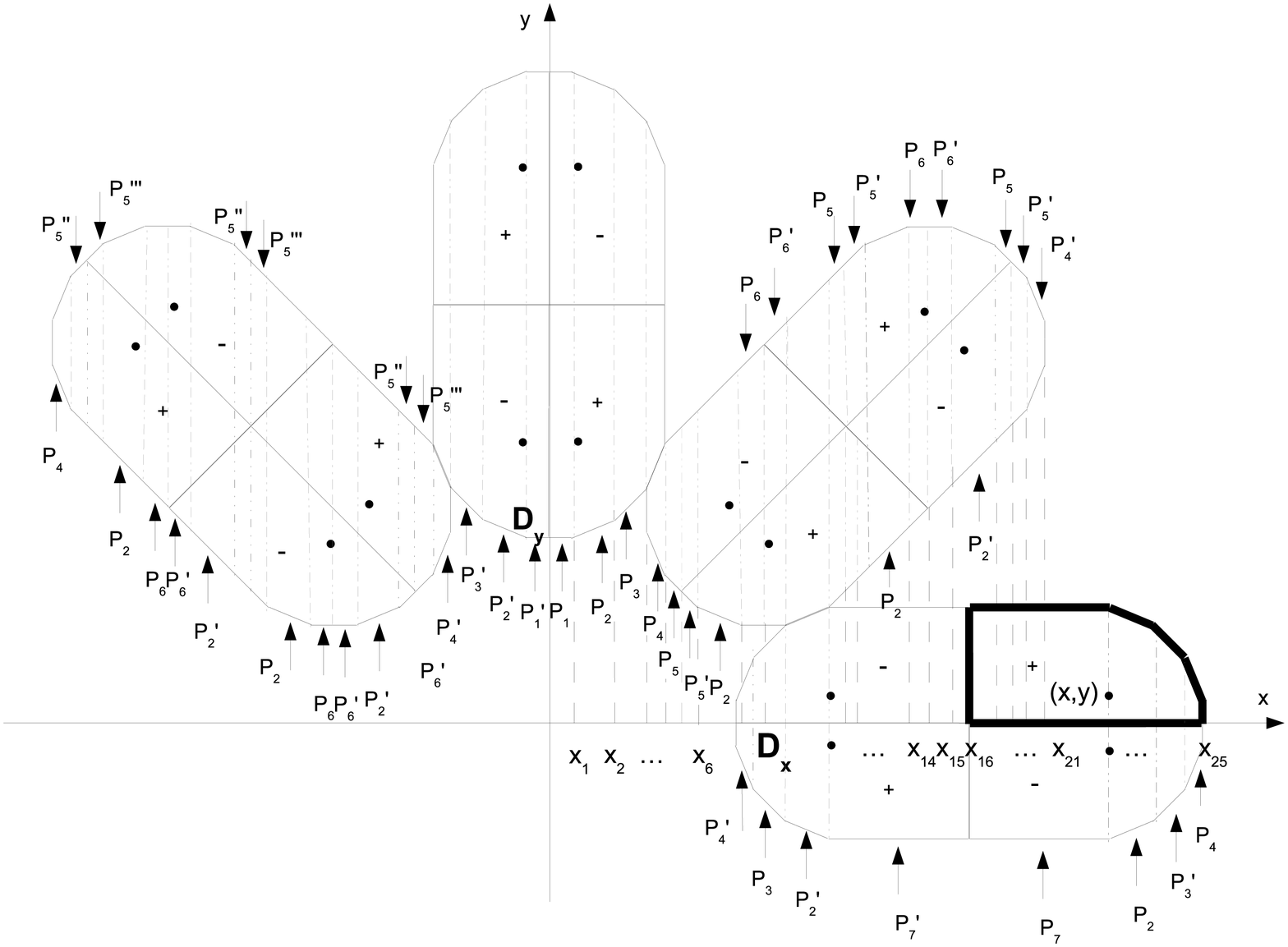,width=10 cm}
\caption{All the fifteen vertical POCs covering completely the EPP of the polygon billiards $A''$. The dashed lines
correspond to respective singular diagonals of the POCs - each two neighbour such diagonals are boundaries of a single POC. POCs with the same index
number have the same periods and the same forms after folding into the billiards (see Fig.9 below). To follow each particular POC one has to apply the
respective periods of Fig.3A}
\end{center}
\end{figure}

For a simplicity we consider the respective quantization on the case $A''$ of the RPBs only.
In this case we choose the direction of the period ${\bf D}_y$, i.e. the vertical one in Fig.3A to which the momentum ${\bf p}$ is parallel. A global
periodic skeleton corresponding to the chosen period consists of the seventeen POCs shown in Fig.8. According to the general rules corresponding to the
case considered we have the following quantization conditions in each POC
\be
{\bf p}\cdot{\bf D}_y=pD_y=8\pi mZ\nn\\
m=0,\pm 1,\pm 2,...
\label{16}
\ee
while the respective basic SWFs has the form \cite{42}
\be
\Psi_{BSWF}^{POC}(x,y)=e^{ipy}(A\sin(\sqrt{2E_0}x)+B\cos(\sqrt{2E_0}x))
\label{17}
\ee
where $p$ and $E_0$ can depend on a POC. However matching $\Psi_{BSWF}^{POC}(x,y)$ defined in each neighbor pair of POCs on their boundaries to get the global
SWF $\Psi_{BSWF}(x,y)$ we find that both $p$ and $E_0$ is the same for all POCs so that the global form of $\Psi_{BSWF}(x,y)$ is again as in \mref{17}. The latter
SWF however has to be periodic on the respective PBRS and in particular by the period ${\bf D}_x/(4Z)$ which gives the following quantization condition
for $E_0$
\be
\sqrt{2E_0}D_x=8\pi Zn\nn\\
n=\pm 1,\pm 2,...
\label{18}
\ee
so that for the respective energy spectrum we have
\be
E_{mn}=\fr{\b p}^2+E_0=8\pi^2Z^2(m^2+n^2)\nn\\
m,n=\pm 1,\pm 2,...
\label{19}
\ee
i.e. the same result as in the aperiodic case.

To get the respective SWF satisfying the Dirichlet boundary conditions we take now $\Psi_{BSWF}(x,y)$ in the form \mref{17} and sum it over all
the points of the EPP of Fig.6 being the mirror reflections of the "initial" point $(x,y)$ of the billiards $A''$. It is easy to check that then all the
contributions from the cosine function in \mref{17} mutually cancel and the remaining form of the global $\Psi_{mn}^{sem}(x,y)$ is exactly the same
(up to a normalization) as in the aperiodic case given by \mref{5}.

\begin{figure}
\begin{center}
\psfig{figure=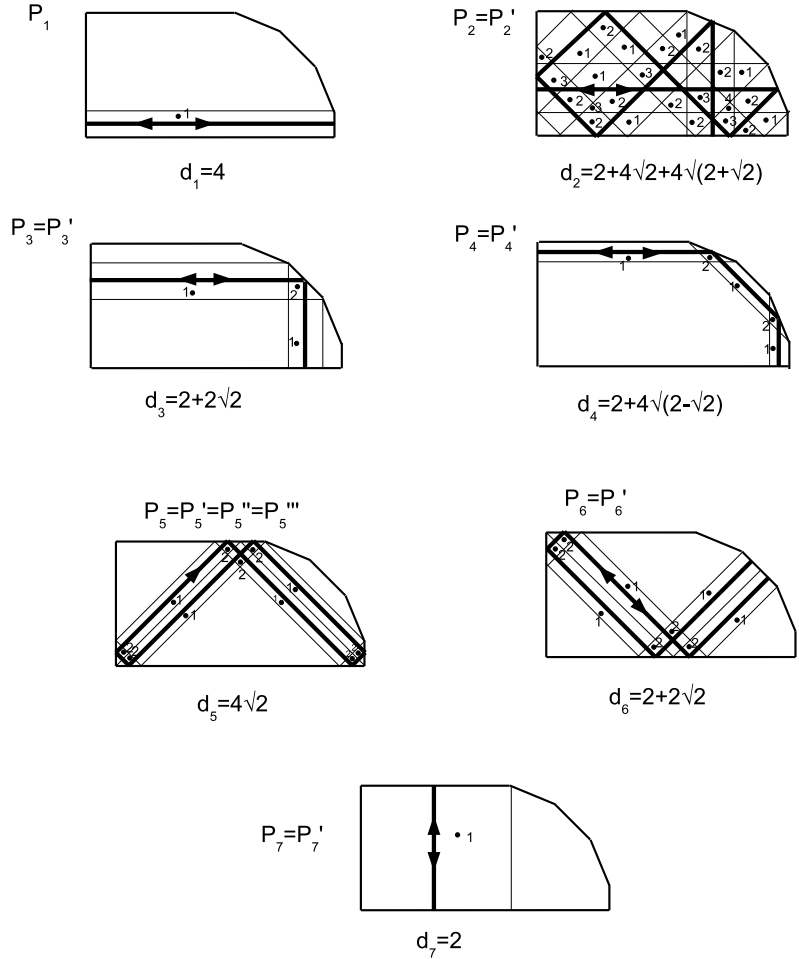,width=12 cm}
\caption{The vertical POCs of the EPP of Fig.8 after folding into the polygon billiards $A''$ with
a distinguished single periodic trajectory (thick lines). The thin lines denote singular diagonals of the respective POCs. The numbers put near the
distinguished billiards points denote multiplicities with which the respective POCs cover the points. The numbers below the billiards figures
are the lengths of the periodic trajectories which each POC is composed of.}
\end{center}
\end{figure}

However by its construction the SWF built on the periodic skeleton is totally composed of contributions from all POCs which the periodic skeleton is
composed of and having the forms of standing waves each.
This fact has been discussed also in our earlier paper \cite{53} where it was shown on several examples of the polygon billiards that the superscar states of
Bogomolny and Schmit \cite{46} built on respective POCs are only components of the global SWFs corresponding to the cases discussed.
The SWF \mref{5} reveals the same property mentioned strengthened by an observation that it vanishes approximately on each singular
diagonal (SD) shown in Fig.6.
This is because according to the figure the equations of the SDs are $x=x_k\;k=0,...,25$, or $x=-x_k,\;k=1,...,21$, with $x_0=0$. But every $x_k$ can
be expressed as a linear combinations of $1$ and the real number $A,B,C$ of the table of App.A.2 with integer coefficients of these combinations.
Therefore making the substitutions $x=x_k=\sum_{l=0}^3x_{kl}X_l,\;X_0=1,X_1=A,X_2=B,X_3=C,\;k=0,...,25,$ in each component of the solution \mref{5} we
get
\be
\left|\Psi_{mn}^{sem}(x_k,y)\right|=\nonumber\\
\left| e^{4\pi mZ(r+2)}\sin\left(4m\pi\sum_{l=1}^3(x_{kl}-\delta_{l2}-\delta_{l3})(ZX_l-q_l)\right)\sin(4n\pi Zy)-\right.\nonumber\\
                     e^{4\pi nZ(r+2)}\sin(4n\pi\left(\sum_{l=1}^3(x_{kl}-\delta_{l2}-\delta_{l3})(ZX_l-q_l)\right)\sin(4m\pi Zy)+\nonumber\\
                     e^{2\pi i(m-n)\sqrt{2}Z(r+2)}\times\nonumber\\
                     \sin\left(2(m-n)\pi\sum_{l=1}^3(x_{kl}'-2\delta_{l2})(ZX_l-q_l)\right)\sin\left(2(m+n)\pi\sqrt{2}Zy\right)-\nonumber\\
                     e^{2\pi i(m+n)\sqrt{2}Z(r+2)}\times\nonumber\\
                     \left.\sin\left(2(m+n)\pi\sum_{l=1}^3(x_{kl}'-2\delta_{l2})(ZX_l-q_l)\right)\sin\left(2(m-n)\pi\sqrt{2}Zy\right)\right|<\nonumber\\
                 4\pi\left((m+n)\sum_{l=1}^3|x_{kl}|+n\sum_{l=1}^3|x_{kl}'|+2m+4n\right)\times N^{-1/3}
\label{20}
\ee
where integers $x_{kl}'$ are defined by the equality $\sum_{l=0}^3x_{kl}'X_l\equiv\sum_{l=0}^3x_{kl}\sqrt{2}X_l$.

Limiting to the area of the polygon billiards $A'$ and $A''$ it means that $\Psi_{mn}^{sem}(x,y)$ vanishes approximately in them
along the lines shown in Fig.Fig.9,10 which are traces of all the SDs of the POCs of Fig.8 when the latter are folded into the billiards. Therefore
this closeness to zero along the lines mentioned is the most visible effect of the periodic structure of the RPRS on which the SWF $\Psi_{mn}^{sem}(x,y)$
is defined showing also that in the area of the billiards it is the coherent interference of standing waves in POCs spanned between their singular diagonals
mimic to some extent the superscar states of Bogomolny and Schmit \cite{46}.
\begin{figure}
\begin{center}
\psfig{figure=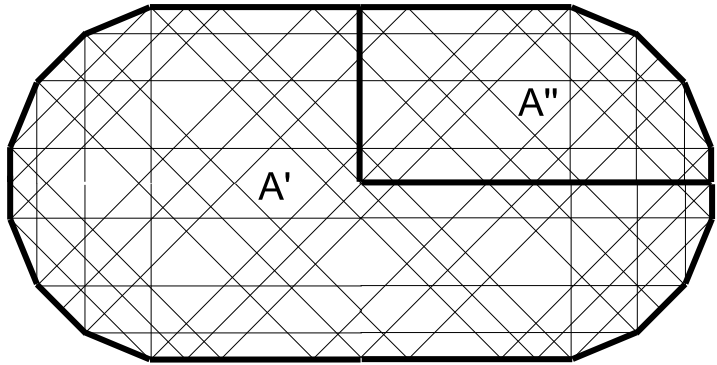,width=10 cm}
\caption{The singular diagonals in the polygon billiards $A'$ and $A''$ on which the SWF \mref{5} approaches approximately zero}
\end{center}
\end{figure}

It is shown also on Fig.9 that the running wave contributions of a particular POC to $\Psi_{mn}^{sem}(x,y)$ depend on a length of a period defining
the POC,
i.e. the longer is the period the more frequent the respective POC cover each billiards point $(x,y)$ and therefore the more frequent the running wave
in a given POC interferes with itself in this point. In the particular example of the SWF considered there are the longest POCs $P_2$ and $P_2'$ each
of which can interfere four times with oneself as it is visible in Fig.7 while the shorter ones $P_3,...,P_6'$ can cover any point at most twice
and the shortest period POCs $P_1$ and $P_7,P_7'$ only once. These notices will be discussed wider in sec.6.

\section{The accuracy of the semiclassical energy spectra for the polygon billiards enveloping the Bunimovich stadia}

\hskip+2em Let us now estimate the accuracy of the semiclassical energy levels defined by \mref{3b} and \mref{12} for the polygon billiards $A'$ when they are
compared with the corresponding levels of the Bunimovich stadium $A$. First we have to estimate the respective accuracy of these formulae for the energy
spectrum of the polygon billiards themselves. We can proceed similarly as in Sec.1 when the respective estimation was done by substituting the Bunimovich
stadia by their polygon envelopes. To this goal we assume further the quantum number to be positive (see the discussion at the end of sec.2.1) and satisfying
$0\leq m<n$ and we calculate distances $l_{mn}(x,y)$ between the curve $L_{mn}$ composed of zeros of
$\Psi_{mn}^{sem}(x,y)$ closest to the polygon boundary and the boundary itself. The distance $l_{mn}(x,y)$ is defined by the vector
${\bf l}_{mn}(x,y)={\bf R}'-{\bf R}=l_{mn}\frac{\bf R}{R}$ shown in Fig.9. Of course $l_{mn}(x,y)$ is equal to zero on
the boundary segments on which $\Psi_{mn}^{sem}(x,y)$ vanishes, i.e. on the segments $a$ and $b$ of the curve $L_{mn}$ on Fig.11. $l_{mn}(x,y)$ can be
estimated on the remaining segments of the polygon boundary using the respective Taylor expansion of $\Psi_{mn}^{sem}(x,y)$.

However as the proper expansion
variable we should take not $l_{mn}(x,y)$ itself but rather the latter divided by a "typical" wave length associated with the SWF $\Psi_{mn}^{sem}(x,y)$.
Taking
into account the quantization conditions \mref{3} or \mref{9} it is seen that as such a "typical" wave length can be taken $((m+n)Z)^{-1}$
(in the billiards length units). Therefore the respective expansion parameter can be the dimensionless variable $(m+n)Zl_{mn}(x,y)$ with the natural
condition $|(m+n)Zl_{mn}(x,y)|<<1$ since it is obvious that $l_{mn}(x,y)$ should be clearly smaller than the wave length $((m+n)Z)^{-1}$.

Therefore assuming that the linear term dominates in the expansion we have
\begin{figure}
\begin{center}
\psfig{figure=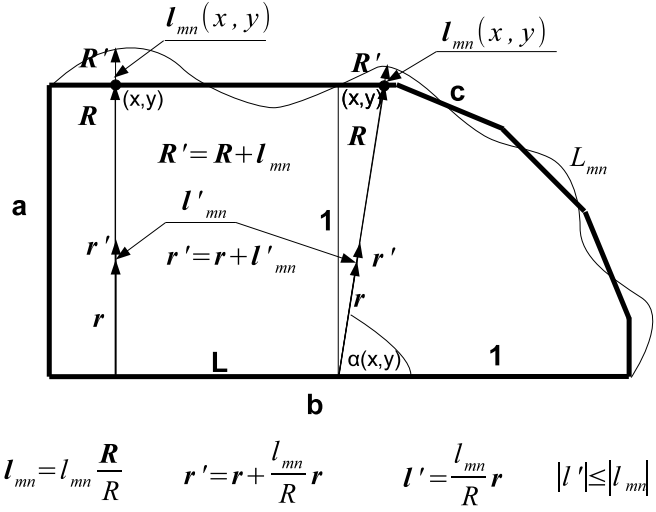,width=10 cm}
\caption{Zeros of $\Psi_{mn}^{sem}(x,y)$ (the thin line $L_{mn}$) closest to the polygon billiards boundary. $L_{mn}$ coincides with the sides {\bf a}
and {\bf b} of the polygon. The figure shows the transformation of the polygon area into the one closed by the curve $L_{mn}$ and
satisfying the conditions of THEOREM 4 of App.C.}
\end{center}
\end{figure}

\be
l_{mn}(x,y)=-\frac{\Psi_{mn}^{sem}(x,y)}{\frac{\bf R}{R}\cdot\bigtriangledown\Psi_{mn}^{sem}(x,y)}
\label{21}
\ee
where $(x,y)$ is any point of the polygon boundary in which $\frac{\bf R}{R}\cdot\bigtriangledown\Psi_{mn}^{sem}(x,y)\neq 0$ as it is assumed.

Taking into account the explicit forms \mref{5} and \mref{13} of the SWFs we see that the gradient action on $\Psi_{mn}^{sem}(x,y)$ multiplies it by
$Z$ and by $m$ and $n$ its different components. Therefore \mref{21} gives us
\be
|(m+n)Zl_{mn}(x,y)|=\ll|\frac{(m+n)Z\Psi_{mn}^{sem}(x,y)}{\frac{\bf R}{R}\cdot\bigtriangledown\Psi_{mn}^{sem}(x,y)}\r|\leq 2\pi\frac{|m|J_x+|n|J_y}{M}\cdot N^{-1/3}
\label{21c}
\ee
where $0<M<32\pi$ is the minimal value of $\ll|\frac{\bf R}{R}\cdot\bigtriangledown\Psi_{mn}^{sem}(x,y)/((m+n)Z)\r|$ on the segment of the polygon boundary on which
the formula \mref{21} is used. To estimate its allowed quantity let us note that in the case considered the calculated $l_{mn}(x,y)$ has to satisfy
additionally the following restriction
\be
|l_{mn}(x,y)|<<
\ll|\frac{2\frac{\bf R}{R}\cdot\bigtriangledown\Psi_{mn}^{sem}(x,y)}{\ll(\frac{\bf R}{R}\cdot\bigtriangledown_{\Psi}\r)^2\Psi_{mn}^{sem}(x,y)}\r|
\label{21a}
\ee
expressing the domination mentioned. In the above formula $\bigtriangledown_{\Psi}$ acts only on $\Psi_{mn}^{sem}(x,y)$.

Using the estimation
$\ll|\Psi_{mn}^{sem}(x,y)\ll(\frac{\bf R}{R}\cdot\bigtriangledown_{\Psi}\r)^2\Psi_{mn}^{sem}(x,y)\r|<32(m+n)^2Z^2\pi(mJ_x+nJ_y)\times N^{-1/3}$ which follows
from \mref{5} and \mref{7} and taking into account \mref{21c} we can rewrite the restriction \mref{21a} as
\be
M>>4\sqrt{\pi(mJ_x+nJ_y)}\times N^{-\frac{1}{6}}
\label{21b}
\ee
defining segments of the billiards boundary where the formula \mref{21} can be applied. On these segments we get finally
\be
|(m+n)Zl_{mn}(x,y)|<<\fr\sqrt{\pi(mJ_x+nJ_y)}\times N^{-\frac{1}{6}}=C_1\times N^{-\frac{1}{6}}
\label{21d}
\ee

Using similar arguments one can convince oneself that including still higher order terms (quadratic, cubic, etc) which can dominate in the
corresponding Taylor expansion we get instead of \mref{21d}
\be
|(m+n)Zl_{mn}(x,y)|<<C_k\times N^{-\frac{1}{3(k+1)}}\nn\\
k=2,3,4,...
\label{22}
\ee

Assuming that in the above possibilities there is a finite number of terms of the Taylor expansion which have to be taken into account we have
\be
|l_{mn}(x,y)|<<C_K((m+n)Z)^{-1}\times N^{-\frac{1}{3(K+1)}}<C_K((m+n)Z)^{-1}=\epsilon_{mn}
\label{23}
\ee
where $K$ is the maximal number of terms which appear in our calculations.

Consider now the curve $L_{mn}$ as the boundary of the domain $D_{mn}$ in which $\Psi_{mn}^{sem}(x,y)$ satisfies the Schrödinger equation (SE)
vanishing on $L_{mn}$. Consider also the exact energy spectrum $E_k^{exact},\;k=1,...,$ to the quarter of the Bunimovitch stadium. The domain $D_{mn}$ defines an energy spectrum
$E_k^{(mn)},\;k=1,...,$ to which belongs the energy $E_{mn}$. According to Theorem 4 of App.C the spectrum $E_k^{(mn)}$
will be close to the exact spectrum $E_k^{exact},\;k=1,...,$ if $L_{mn}$ is sufficiently close to the quarter of the Bunimovich stadium boundary
and this quarter can be continuously and vanishingly transformed into the domain $D_{mn}$. However it is easy to note that such a transformation is given
as the composition of the one given by ${\bf l}_{mn}'(x,y)$ of Fig.11 and the one ${\bf l}_{mn}^{pol}(x,y)$ given by \mref{2}, i.e. we have
\be
{\bf l}_{mn}^{com}(x,y)={\bf l}_{mn}^{pol}(x,y)+{\bf l}_{mn}'(x+l_{mn,x}^{pol}(x,y),y+l_{mn,y}^{pol}(x,y))
\label{24}
\ee
where ${\bf l}_{mn}^{com}(x,y)$ denotes shifting of the point $(x,y)$ of the quarter of the Bunimovitch stadium transforming it into the domain $D_{mn}$.

According to \mref{2b} and \mref{23} we have therefore
\be
|{\bf l}_{mn}^{com}(x,y)|<\epsilon_{pol}+\epsilon_{mn}
\label{25}
\ee
and according to Theorem 4 of App.C the energies $E_k^{(mn)}$ approximate $E_k^{exact}$ by
\be
\ll|\frac{E_k^{(mn)}}{E_k^{exact}}-1\r|<\eta_{mn}
\label{26}
\ee
for some $\eta_{mn}$ and for each pair $m,n$ satisfying the inequality $mJ_x+nJ_y<<N^{1/3}$.

Of course among all $k,\;k=1,...$, there are $k_{mn}$ for which we have
\be
\ll|\frac{E_{mn}}{E_{k_{mn}}^{exact}}-1\r|<\eta_{mn}\nn\\
mJ_x+nJ_y<<N^{1/3}
\label{27}
\ee

Taking therefore the maximal $\eta_{max}$ from the set $\{\eta_{mn}:\;mJ_x+nJ_y<<N^{1/3}\}$ we get
\be
\ll|\frac{E_{mn}}{E_{k_{mn}}^{exact}}-1\r|<\eta_{max}
\label{28}
\ee
for each pair $m,n$ satisfying $mJ_x+nJ_y<<N^{1/3}$.

Noticing yet that since $\epsilon_{pol}\approx \epsilon_{max}$, one can expect that the accuracy given by \mref{28} is determined
almost equally by the polygon enveloping of the Bunimovich stadium and by the respective semiclassical approximation.

\begin{figure}
\begin{center}
\psfig{figure=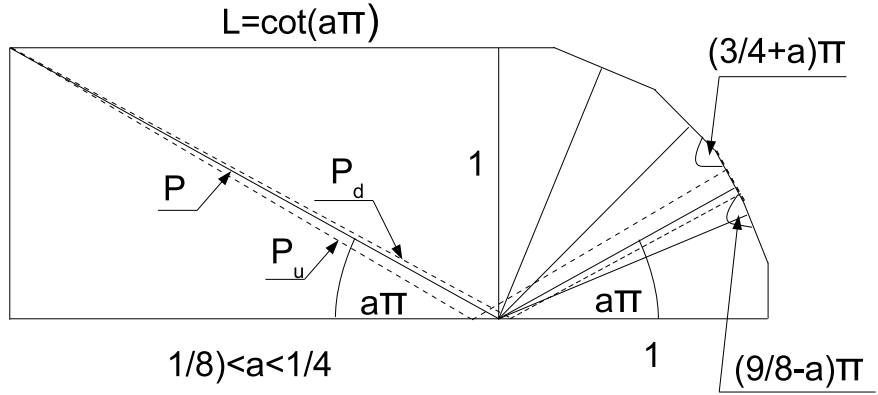,width=13 cm}
\caption{The Bunimovich stadium enveloped by a polygon billiards with an irrational $a$. The periodic trajectory $P$ of the original
billiards is substituted by one of the two others $P_d$ or $P_u$ depending on the rational approximation of $a$, i.e whether an
approximating rational is larger or smaller than $a$ respectively.}
\end{center}
\end{figure}

\section{Enveloping the Bunimovich stadium by an irrational polygon billiards}

\hskip+2em Essentially our considerations of the case, see Fig.12, can be reduced to the previous ones since approximating an irrational $a$ by a
sufficiently accurate rational the geometrical organization of the shortest periods of the original Bunimovich billiards is changed only slightly and can be done
arbitrarily small. Since also a rational polygon billiards obtained in this way approximate the irrational one with an arbitrary good accuracy the
results of the discussion on the beginning of sec.2 remains valid as well. Serious changes can be expected mostly in the form of the corresponding EPP due
to a denominator $D$ of a rational approximating $a$. Namely if $D$ appears to be large then the common least multiple corresponding to it and to the
remaining rational angles of the RPB considered has to be also large as well as a genus of the respective multitorus, i.e. a number of independent
periods can become large in comparison with any of the cases considered earlier. This further can complicate detailed considerations of such
rational approximation of the irrational case. Nevertheless the procedure applied in the previous sections to construct SWFs and the respective energy
spectra can be repeated also in principle here. In particular relations between the periods of the rationalized polygon billiards can be analyzed in a way similar
to the one of the previous sections, i.e. the coefficients $a_q^{ijk\to rst},\;q=x,y$, in \mref{1} can be expressed by a linear combinations of
some number of independent irrationals $X_k,\;k=1,...,n$, similarly to \mref{3e}. This conclusion follows from the fact that these coefficients can be
obtained as in the cases already considered, i.e. by projections of a respective number of sides of the considered rational polygon which the sides treated
as vectors can represent each period of the respective EPP by their sum. Both their lengths as well as their projections on the two chosen period
directions are given by trigonometric functions of rational angles and their products. Among these functions one can always select a finite number of
them forming an algebra with rational coefficients.

\section{Summary and conclusions}

\hskip+2em In this paper we have shown that it is possible to construct for some part of the antisymmetric high energy spectrum of the Bunimovich stadium
and for the respective antisymmetric wave functions their semiclassical approximations. These approximations are obtained by the procedure having the
following basic properties
\begin{enumerate}
\item it approximates the Bunimovich stadium by its polygon envelopes;
\item it uses a number of shortest periodic orbits of the Bunimovich stadium to built its polygon envelopes;
\item a number of periodic orbits used to built respective envelopes is not limited;
\item polygon envelopes obtained by the procedure include all the periodic orbits of the Bunimovich stadium used to build them;
\item SWFs built on the polygon envelopes of the Bunimovich stadium have properties typical for the polygon billiards, i.e. the superscar structure
\cite{53}, by which SWFs built for the cases considered vanish on the superscar diagonals;
\item the accuracy of the constructed SWFs describing the distinguished parts of the Bunimovich stadia energy spectra is controlled by the respective
theorems of App.C;
\item while the considered cases of the Bunimovich stadia have been enveloped by the rational polygon billiards the cases when the Bunimovich stadium enforces its enveloping by an irrational polygon billiards can be considered by substituting first
the latter billiards by its rational approximation;
\item by its nature the method used to describe semiclassically quantum states in the Bunimovich stadia covers the high energy
regions of their spectra - the higher the more precise this description is to be;
\item the form \mref{3b} of the energy spectra provided by the used method of the semiclassical approximation shows that it selects from the whole
spectra only those levels which can be arranged in a pattern typical for the rectangular billiards;
\end{enumerate}

Considering the superscars structure of the constructed SWFs one can expect that including more and more periodic orbits of the Bunimovich stadium to
construct its polygon envelopes one gets the scars structures of the limit wave functions identified by Heller \cite{44}. The following
notes can suggest that such expectations may be real despite the fact that energies considered by Heller are much smaller than ones
considered in our paper.

\begin{itemize}
\item As it was noticed in the point 8. above including still more periodic orbits shifts the energy spectra still to higher regions;
\item in our high energy semiclassical description of the quantum states in the Bunimovich stadium the periodic structure of the classical motion in
the billiards manifests itself by vanishing of the respective SWFs on the POC diagonals;
\item the more isolated orbits are used to build a polygon envelope the closer to them POC's diagonals are running, i.e. the orbits are pinched
by the respective diagonals;
\item while widths of POCs defined by isolated and unstable periodic orbits decreases with the growing number of the orbits
a number of the wave lengths between POC's diagonals grows rapidly in such cases and is of order $Z$ - such
a growing can therefore generate an effect of amplification of mean values of SWFs between POC's diagonals;
\item there are POCs (see for example $P_6$ in Fig.9) in which one of their diagonals permanently occupies a limiting position of a periodic orbit defining
the POCs when a number of included periods grows infinitely - it suggests that in the positions of the respective periodic orbits one can observe a nodal
line of the exact wave function rather than its amplified amplitude, i.e. rather an anti-scar than a scar;
\end{itemize}

Extrapolating therefore runnings of POCs with short and long isolated periodic orbits shown in Fig.Fig.7,8 through the billiards one can expect the
following picture when a number of included periodic orbits grows

\begin{itemize}
\item the shortest periodic orbits can manifest themselves as scars of SWFs or as anti-scars (nodal lines) independently of energy;
\item the longest periodic orbits should be transformed into the chaotic background of the high energy wave functions.
\end {itemize}

The latter conclusion can be justified noticing that the longer is a periodic orbit the more frequently its POC crosses vicinities of any
point of the billiards and its directions in these vicinities become "chaotic".

\appendix

\section{Linear relations on the plane between the periods in the cases $A'',\;B''$ and $C''$ of the polygon billiards shown in Fig.1}

\subsection{The case $A''$}

\hskip+2em Considering the geometry of the EPP of Fig.3A we can note that each period linking a pair of two parallel sides of the EPP can be represented
as a sum of its respective sides considered as vectors and the vectors ${\bf r}_i,\;i=1,...,4$ of Fig.3A. For example according to Fig.3A the period
${\bf D}_{314\to 444}={\bf r}_4-{\bf r}_3+\vec{445}+\vec{444}-\vec{345}-\vec{344}-\vec{343}-\vec{342}-\vec{341}-\vec{311}-\vec{312}-\vec{313}$. If
projected on the $x,y$ axes the lengths of these vectors are always multiplied by $\pm\sqrt{2}/2=\pm A/2,\; \pm\sin(\pi/8)=\pm C/2$ or $\pm\cos(\pi/8)=\pm B/2$.
Since $\ll|{\bf r}_k\r|=r=B+C-1,\;k=1,...,4,\;|\vec{ij5}|=d/2=-1-A+B+C$ and $|\vec{ijk}|=d,\;k\neq 5$, then it follows from Table \ref{table:algebra} below that ${\bf D}_{314\to 444}$
is a linear combination of ${\bf D}_x={\bf D}_{311\to 321}=2[1,0]$ and ${\bf D}_y={\bf D}_{121\to 111}=2[0,1]$ with coefficients which are also
linear combinations of the following four real numbers $X_0=1,X_1=A,X_2=B,X_3=C$ with rational coefficients which denominators are not larger than $4$,
i.e. we can write
\be
{\bf D}_{314\to 444}=a_x^{314\to 444}{\bf D}_x+a_y^{314\to 444}{\bf D}_y\nn\\
a_k^{314\to 444}=\frac{1}{4}\sum_{i=0}^3a_{ki}^{314\to 444}X_i,\;\;\;k=x,y\nn\\
X_0=1,\;X_1=A=\sqrt{2},\;X_2=B=\sqrt{2+\sqrt{2}},\;X_3=C=\sqrt{2-\sqrt{2}}
\label{A1}
\ee
and $a_{ki}^{314\to 444}$ are all integer.

\subsection{The cases $B''$ and $C''$}

\hskip+2em The main difference between the present cases and the previous one lies in the growing complexity of the EPPs of the former in comparison
with the EPP of the case $A''$ as it can be seen from the figures 3, 4 and 5 and expresses in others four irrational numbers determining the
coefficients of the linear relations on the plane between the independent periods of both the cases. Namely, taking as previously for each case the
periods  ${\bf D}_x=2[1,0]$ and ${\bf D}_y=2[0,1]$ as the base on the $x,y$-plane we get for any period ${\bf D}_{ijk\to rst}$
\be
{\bf D}_{ilk\to rst}=a_x^{ilk\to rst}{\bf D}_x+a_y^{ilk\to rst}{\bf D}_y\nn\\
a_k^{ilk\to rst}=\frac{1}{4}\sum_{q=0}^7a_{kq}^{ilk\to rst}X_q,\;\;\;k=x,y\nn\\
X_0=1,\;X_1=A,\;X_2=B,\;X_3=C,\;
X_4=D=\sqrt{2+\sqrt{2+\sqrt{2}}},\nn\\
X_5=E=\sqrt{2+\sqrt{2-\sqrt{2}}},\;
X_6=F=\sqrt{2-\sqrt{2+\sqrt{2}}},\;X_7=G=\sqrt{2-\sqrt{2-\sqrt{2}}}
\label{A2}
\ee
and $a_{kq}^{ijk\to rst}$ are all integer.

The latter coefficients are obtained by the polygon side projections of the polygons composing the EPPs
of Fig.4 and Fig.5 on the $x,y$-axes. The projected sides are then always multiplied by one of the following numbers
\be
\pm\cos\frac{\pi}{4}=\pm\sin\frac{\pi}{4}=\pm\fr A,\;\;\pm\cos\frac{\pi}{8}=\pm\fr B,\;\;\pm\sin\frac{\pi}{8}=\pm\fr C,\nn\\
\pm\cos\frac{\pi}{16}=\pm\fr D,\;\;\pm\sin\frac{\pi}{16}=\pm\fr F,\;\;
\pm\cos\frac{3\pi}{16}=\pm\fr E,\;\;\pm\sin\frac{3\pi}{16}=\pm\fr G
\label{A3}
\ee

The irrationals $X_q,\;q=1,...,7$, together with $X_0=1$ form a linear eight dimensional algebra with the multiplication rules given by the Table 1 below.
The rules of the table follow from \mref{A3} and from the well known trigonometric identities such as
\be
\sin\alpha\cos\beta=\fr(\sin(\alpha+\beta)+\sin(\alpha-\beta))
\label{A5}
\ee
and similar.

\begin{table}[ht]
\caption{Product algebra of $X_q$}
\centering
\begin{tabular}{c|ccccccc}
\hline\hline
&A&B&C&D&E&F&G\\
\hline
A&2&B+C&B-C&E+G&D+F&E-G&D-F\\
B&&2+A&A&D+E&D+G&-F+G&E+F\\
C&&&2-A&F+G&E-F&D-E&D-G\\
D&&&&2+B&A+B&C&A+C\\
E&&&&&2+C&A-C&B\\
F&&&&&&2-B&-A+B\\
G&&&&&&&2-C
\end{tabular}
\label{table:algebra}
\end{table}

The reciprocal elements of $A,...,G$ are following
\be
A^{-1}=\fr A,\;\;B^{-1}=\fr ABC,\;\;C^{-1}=\fr AB,\;\;D^{-1}=\fr (2+A)(2-B)D,\nn\\
E^{-1}=\fr (2-A)(2-C)E,\;\;F^{-1}=\fr (2+A)(2+B)F,\;\;G^{-1}=\fr (2-A)(2+C)G
\label{A4}
\ee

The above relations allow us altogether to express all the coefficients of the relations \mref{A1} as a linear combinations of the
elements $1,A,...,G$ with integer coefficients.

\section{The Dirichlet simultaneous approximation theorem \cite{1}}

\begin{tw}
For any real numbers $X_1,...,X_n$ and any natural $N$ there exist integers $q_1,...,q_n$ and $0<C\leq N$ which satisfy the condition
\be
|CX_k-q_k|<\frac{1}{N^\frac{1}{n}},\;\;\;\;\;\;\;   0<C\leq N\nn\\
k=1,...,n
\label{A7}
\ee
\end{tw}

\section{Smooth behavior of energy levels as a function of a billiard boundary - general theorems}

\hskip+2em Consider two billiards which are close to each other in the meaning of the following theorem proved in the monography of Courant and Hilbert
\cite{40}.

\begin{de} It is said that the domain $G$ is approximated by the domain $G'$ with the $\epsilon$-accuracy if $G$ together with its
boundary can be transformed pointwise into the domain $G'$ together with its boundary by the equations
\be
x'=x+g(x,y)\nn\\
y'=y+h(x,y)
\label{A29}
\ee
where $g(x,y),\;h(x,y)$ are both piecewise continuous and less in $G$ in their absolute values than a small positive number $\epsilon$ together with
their first derivatives.
\end{de}

\begin{de}
If all conditions of Definition 1 are satisfied while $\epsilon\to 0$ then it is said that $G$ is a continuous deformation of $G'$.
\end{de}

\begin{tw}
Let $G$ and $G'$ satisfy all conditions of Definition 1. Then for any boundary condition $\p\Psi/\p n+\sigma\Psi=0$ the energy spectrum
corresponding to $G'$ approximates the one of $G$ with the $\epsilon$-accuracy. More precisely for any $\epsilon$ there is a number $\eta$
depending only on $\epsilon$ and vanishing with it such that for respectively ordered energy levels $E'_n$ and $E_n$ corresponding to the domains
$G'$ and $G$ we have
\be
\ll|\frac{E_n'}{E_n}-1\r|<\eta
\label{A30}
\ee
\end{tw}

\begin{tw}
Let $G$ and $G'$  satisfy the conditions of Theorem 2  and $G$ is a continuous deformation of $G'$ then the energy spectrum
corresponding to $G'$ varies continuously with $\epsilon\to 0$ approaching the energy spectrum of $G$ controlled by the conditions \mref{A30}.
\end{tw}

\begin{tw}
Theorem 3 remains valid with none condition on the first derivatives of $g(x,y),\;h(x,y)$ in the case of the Dirichlet boundary condition $\Psi=0$.
\end{tw}

\begin{tw}
If $G$ and $G'$ are transformed each into other by \mref{A29} and the absolute value of the Jacobean of the latter transformation is bounded from above
and below than the ratio $E_n'/E_n$ for respectively ordered energy levels $E'_n$ and $E_n$ corresponding to the domains $G'$ and $G$ satisfy for
sufficiently large $n$ the following relation
\be
0<a<\ll|\frac{E_n'}{E_n}\r|<b
\label{A31}
\ee
where $a$ and $b$ are independent of $n$.
\end{tw}

\hskip+2em


\begin{thebibliography}{99}
\bibitem{3}Gutzwiller M. C., {\it "Chaos in Classical and Quantum   Mechanics"} (New York: Springer 1990)
\bibitem{4} Maslov V.I. and Fedoriuk M.V., {\it Semi-classical Approximation
in Quantum Mechanics} (Dordrecht, Boston, London: Reidel 1981)
\bibitem{41}Stefan Giller, Jaros{\l}aw Janiak,  {\it Acta Phys. Pol.} {\bf B 44} (2013) 1725-1764
\bibitem{42}Stefan Giller, {\it Acta Phys. Pol.} {\bf B 46} (2015) 801-842
\bibitem{53}Stefan Giller, {\it J. Mat. Phys.} {\bf 59} 072107 (2018)
\bibitem{46}Bogomolny E. and Schmit C., {\it Phys. Rev. Lett.} {\bf 92} (2004) 244102
\bibitem{44}Heller, E.J., {\it Phys. Rev. Lett.} {\bf 53}, (1984) 1515
\bibitem{40} Courant R. and Hilbert D., {\it Methods of Mathematical Physics}, p.421, (NY, London: Intercience Publishers 1953)
\bibitem{5}Arnold V.I., {\it Mathematical Methods of Classical Mechanics} (Berlin: Springer Verlag 1978)
\bibitem{52}Bunimovich L.A. {\it Common. Math. Phys.} {\bf 65} (1979) 295-312
\bibitem{1}J.W.S. Cassels, {\it "An introduction to diophantine approximation"} (Cambridge Univ. Press 1957)
\end{thebibliography}
\end{document}